\documentclass[a4paper, 10pt, oneside,superscriptaddress]{revtex4-1}
\pdfoutput=1 

\usepackage[latin1]{inputenc}
\usepackage{amssymb}
\usepackage{amsmath}
\usepackage{amsthm}
\usepackage{bbold}
\usepackage{braket}
\usepackage{graphicx}
\usepackage{simplewick}
\usepackage[toc,page]{appendix}
\usepackage{enumerate}

\usepackage[usenames,dvipsnames]{color}
\usepackage{hyperref}
\hypersetup{
  colorlinks=true,
 linkcolor=blue,
 citecolor=red}

\newcommand{\be}{\begin{equation}}
\newcommand{\ee}{\end{equation}}

\begin{document}
\title{Quasi locality of the GGE in interacting-to-free quenches in relativistic field theories}

\author{Alvise Bastianello} 
\affiliation{SISSA, via Bonomea 265, 34136 Trieste, Italy}
\affiliation{INFN, Sezione di Trieste, Italy}

\author{Spyros Sotiriadis}
\affiliation{Institut de Math\'{e}matiques de Marseille, (I2M) Aix Marseille Universit\'{e}, CNRS, 
Centrale Marseille, UMR 7373, 39, rue F. Joliot Curie, 13453, Marseille, France }
\affiliation{University of Roma Tre, Department of Mathematics and Physics, 
L.go S. L. Murialdo 1, 00146 Roma, Italy}
\date{\today}

\begin{abstract}

We study the quench dynamics in continuous relativistic quantum field theory, more specifically the locality properties of the large time stationary state. After a quantum quench in a one-dimensional integrable model, the expectation values of local observables are expected to relax to a Generalized Gibbs Ensemble (GGE), constructed out of the conserved charges of the model. Quenching to a free bosonic theory, it has been shown that the system indeed relaxes to a GGE described by the momentum mode occupation numbers. We first address the question whether the latter can be written directly in terms of local charges and we find that, in contrast to the lattice case, this is not possible in continuous field theories. We then investigate the less stringent requirement of the existence of a sequence of truncated local GGEs that converges to the correct steady state, in the sense of the expectation values of the local observables. While we show that such a sequence indeed exists, in order to unequivocally determine the so-defined GGE, we find that information about the expectation value of the recently discovered quasi-local charges is in the end necessary, the latter being the suitable generalization of the local charges while passing from the lattice to the continuum. Lastly, we study the locality properties of the GGE and show that the latter is completely determined by the knowledge of the expectation value of a countable set of suitably defined quasi-local charges.

\end{abstract}

\maketitle

\section{Introduction}

A problem of central interest in current physics research is the study of quantum many-body systems out of equilibrium \cite{rew2,s-jstat1,s-jstat2,s-jstat3,s-jstat4,s-jstat5,s-jstat6,s-jstat7,s-jstat8} which recent experimental progress has made accessible in the laboratory \cite{s-jstat7,exp1,exp2,exp3,exp4,exp5}. Of particular interest is the problem of \emph{quantum quenches} \cite{cft2,cft1,cft3}, in which some parameter of the Hamiltonian of an isolated quantum system is abruptly changed bringing the system in an out-of-equilibrium state. In this protocol the system is initialized typically in the ground state of the pre-quench Hamiltonian and let to evolve following the post-quench Hamiltonian. Even though the evolution of a closed quantum system is unitary (non-dissipative), in the thermodynamic limit the expectation values of local observables exhibit relaxation to a steady state. In a general system, relaxation is expected to lead to a thermal ensemble \cite{out1,out2,out3}, but in the special class of \emph{quantum integrable models} \cite{integrability1,integrability2} this is not true. Even though the expectation value of local observables exhibits relaxation \cite{out4,out5,out6,out7,out8,out8b,out8c,out9,int1,int2,int3,int4,int5,int6,int7,int8,int9,int10,int11,int12,int13,int14,int15,int16,int17,int18,int19,intf1,intf2,intf3,intf4}, the steady state is not described by a thermal ensemble, but rather by a \emph{Generalized Gibbs Ensemble} (GGE) \cite{int1} that retains much more information about the initial state than the conventional Gibbs Ensemble which keeps only memory of the total energy. The fact that a large amount of information about the initial conditions is conserved through the time evolution is due to the defining property of integrable models, that possess an infinite number of independent \emph{local conserved charges} $\mathcal{I}_{j}$. The relaxation of local observables to a suitable GGE density matrix $\rho_{GGE}$ has been confirmed by many different studies \cite{int1,int2,int3,int4,int5,int6,int7,int8,int9,int10,int11,int12,int13,int14,int15,int16,int17,int18,int19,intf1,intf2,intf3,intf4}. 

More explicitly, the statement is that
\begin{equation}
\lim_{t\to\infty}\braket{\prod_{j=1}^{n}\mathcal{O}_{j}^{H}(x_{j},t)}{}_{0}=\text{Tr}\left[\prod_{j=1}^{n}\mathcal{O}_{j}(x_{j})\rho_{GGE}\right]\,\,\, ,
\end{equation}
where on the left hand side, the expectation value $\braket{...}_{0}$ is taken on the initial state and the local observables $\mathcal{O}_{j}^H$ are evolved in the Heisenberg representation up to time $t$. 

In the original proposal and in early studies of the GGE, its density matrix was defined to include as constraints all \emph{local} conserved charges $\mathcal{I}_{j}$: 
\begin{equation}
\rho_{LGGE}= \frac{1}{\mathcal{Z}}e^{-\sum_{j=0}^\infty\lambda_{j}\mathcal{I}^{\text{L}}_{j}}\,\,\, .\label{localGGE}
\end{equation}
We will call this version \emph{local GGE} (LGGE) to distinguish it from other versions introduced below. The Lagrange multipliers $\lambda_{j}$ are fixed in such a way that the expectation values of the charges predicted by the GGE are the same as the values at the initial time $\braket{\mathcal{I}^\text{L}_{j}}_{0}=\braket{\mathcal{I}^\text{L}_{j}}_{GGE}$.

In order to clarify the subsequent discussion, we must further comment on the definition of the local GGE (\ref{localGGE}): in fact, we can give two different interpretations of it.
In particular, we introduce the \emph{direct definition} of the local GGE (DLGGE) as follows: the LGGE is defined as exponential of an operator, the latter unambiguously expressed in terms of the local charges. This amounts to firstly fixing the set of the Lagrange multipliers $\{\lambda_j\}_{j=0}^\infty$, then constructing with them the operator $\sum_{j=0}^\infty \lambda_j \mathcal{I}_j^\text{L}$ and finally exponentiating it in order to define the density matrix (\ref{localGGE}): of course, the Lagrange multipliers must be chosen to obtain the correct expectation value of the local charges.
In the following, we will say that the DLGGE correctly describes the steady state if for any local observable and their correlators the following relation holds
\begin{equation}
\lim_{t\to\infty}\braket{\prod_{l=1}^{n}\mathcal{O}_{l}^{H}(x_{l},t)}{}_{0} = \mathrm{Tr} \left(\rho_\text{DLGGE} \prod_{l=1}^{n}\mathcal{O}_{l}(x_{l})\right)\,\,\, ,\label{DLGGE}
\end{equation}
regardless of the positions of the local observables. 

Closely related to the previous definition, but in a somewhat more constructive perspective, we introduce the notion of \emph{convergence of the truncated LGGE} \cite{int16} (TLGGE) where a sequence of LGGEs with a finite number of charges is constructed
\be
\rho_{TLGGE}^{(N)}=\frac{1}{\mathcal{Z}_N}e^{-\sum_{j=0}^N\lambda_{j}^{(N)}\mathcal{I}_j^\text{L}}
\ee
and we will say that the TLGGE correctly describes the steady state if there exists a sequence of truncated LGGEs such that
\begin{equation}
\lim_{t\to\infty}\braket{\prod_{l=1}^{n}\mathcal{O}_{l}^{H}(x_{l},t)}{}_{0} = \lim_{N\to\infty}\mathrm{Tr} \left(\rho_\text{TLGGE}^{(N)} \prod_{l=1}^{n}\mathcal{O}_{l}(x_{l})\right)\,\,\, ,\label{TLGGE}
\end{equation}
for any correlator of local operators. Both the above notions of LGGE are well posed questions, however the second is clearly a less stringent requirement, while still perfectly meaningful on a physical ground.
Even though on the lattice \cite{int16} the two notions (\ref{DLGGE}) and (\ref{TLGGE}) are essentially equivalent, in the continuous model discussed in the present work it is important to keep them as separate issues and investigate both of them.

Early verifications of the GGE in essentially non-interacting systems focused on the \emph{momentum-space} version of the GGE (MGGE)
\begin{equation}
\rho_{MGGE}= \frac{1}{\mathcal{Z}}e^{-\sum_{k}\lambda(k)E(k) a^\dagger_k a_k}\,\,\, ,\label{momentumGGE}
\end{equation}
which was typically expected to be equivalent to the local GGE in terms of its direct definition (\ref{DLGGE}). In fact this is generally true for non-interacting systems on a lattice. More recently however, it was discovered that in genuinely interacting systems and in the presence of bound states, the GGE as defined by (\ref{localGGE}) i.e. including only local conserved charges, gives incorrect results for general initial states \cite{qlc3,fail1,fail2} and a larger class of conserved charges, the set of \emph{quasi-local charges}, must also be included in order for this \emph{complete or quasilocal GGE} (QLGGE) to give the correct description of the steady state \cite{quenchaction,qlc1,qlc2,qlc4,qlc5,qlc6,qlc7,Panfil,Doyon, s-jstat8,PME2}. 
\begin{equation}
\rho_{QLGGE}= \frac{1}{\mathcal{Z}}e^{-\sum_{j}\lambda_{j}\mathcal{I}^{\text{QL}}_{j}}\,\,\, .\label{QlocalGGE}
\end{equation}
In principle, even in this case we should distinguish between the more direct definition and the notion of the convergence of the truncated GGE, however in the following we will always implicitly refer to the direct definition of QLGGE.

Local charges (sometimes called \emph{ultra-local}) are defined as spatial sums (integrals, in the continuum) of local charge-densities, i.e. operators with spatial support within a finite number of neighbouring lattice sites (operators involving only a finite order of spatial derivatives at a single point, in the continuum). 
In lattice systems milder forms of locality have been considered arriving at the definitions of \emph{pseudolocal} and \emph{quasilocal} charges \cite{Doyon, s-jstat8}. Pseudolocal charges are defined through extensivity of their operatorial norm and include as special cases the quasilocal and the local charges.
Locality properties of the quasilocal charges lay in between the weakest definition of pseudolocality and the strongest concept of locality: like local charges, they can be written as a sum on the lattice of a quasilocal density. Quasilocal densities, in contrast with the local ones, can couple lattice sites infinitely far apart, but the strength of this coupling is exponentially damped with the distance.

While a \emph{complete GGE} \cite{qlc6} that includes quasi-local charges has by now been established as the correct description of the steady state in lattice and spin chain systems, the situation is less clear in continuous integrable systems. Quasi-local charges for continuous quantum field theories were first introduced in \cite{Panfil} and it was argued that not all physically meaningful initial states would relax to a GGE that can be expressed in terms of local charges. Instead quasi-local charges may be needed in general. Moreover it had been already shown earlier that local charges may be ill-defined (UV divergent) for some very natural choices of initial states \cite{caux-denardis, int18}. 

In order to test the LGGE protocol in continuous models, here we will focus on interacting-to-free quantum quenches: in particular we will assume that the evolution follows the simplest relativistic massive bosonic field theory, the Klein-Gordon theory. Despite the absence of interactions during the evolution, in fact the presence of arbitrary interactions in the initial state makes the problem non-trivial. 
In \cite{intf3} it has been shown that for any quench of the above type whose initial state satisfies the cluster decomposition principle, the corresponding steady state is always given by the momentum-space form (\ref{momentumGGE}) (see also \cite{SotiriadisMartelloni,Sotiriadis,BS16} for extensions). Clustering of correlations is a general feature of ground states of local quantum field theories, with massive theories satisfying exponential clustering, while massless ones satisfying weaker (algebraic) clustering \cite{Weinberg}. Its crucial role for the validity of Gaussian relaxation had been already pointed out in analogous studies of interacting-to-free quenches in bosonic lattice models \cite{out8,out8b} and more recently in fermionic ones \cite{out8c}.

In the present work we firstly investigate the direct definition of local GGE (Section \ref{massquench} and \ref{mainsec}), i.e. if the exact steady state of the form (\ref{momentumGGE}) can be recast into the ultra-local form as frequently assumed. Already by studying the simplest possible example, a mass quench, we will demonstrate that the two forms are not equivalent. By comparing the Klein-Gordon model with its lattice version, i.e. a system of harmonic oscillators with short-range couplings, we realize that the above described discrepancy disappears: the lattice version of the local GGE in the strongest interpretation (\ref{DLGGE}) is indeed equivalent to the corresponding lattice version of (\ref{momentumGGE}). 
We will then show in full generality that the failure of (\ref{DLGGE}) in the continuum is not an accidental peculiarity of this special type of quench, but it is always present for any choice of the pre-quench interaction. Our proof uses the K\"all\'en-Lehmann spectral decomposition of the initial state and relies on the observation that the ultra-local GGE written in momentum space imposes analyticity requirements on $\lambda(k)$ that are inconsistent with the correct form.
More importantly we will show that their difference is essential and physically relevant in the sense that the DLGGE gives incorrect predictions for the two point correlation function in the steady state, not only at finite but even at large distances. We emphasize that a basic qualitative characteristic like the long distance correlation function is precisely what a field theoretic approach is supposed to capture. 

In Section \ref{truncGGE} we then address the validity of the local GGE in its weaker sense, i.e. the convergence of its truncated version. In particular, we investigate this issue at two levels: the first question we pose is if there exists a sequence of TLGGEs that guarantees the validity of (\ref{TLGGE}). Indeed, we find a positive answer: we explicitly construct a suitable sequence that reproduces the correlators of local observables in the steady state. 
The second question we investigate concerns how the aforementioned sequence can be determined from the knowledge of the expectation value of suitable conserved charges: the question is delicate, since the expectation values of the local charges turn out to be UV divergent. 
We circumvent this problem by considering the quasi-local charges, in particular we show that the knowledge of the expectation value of charges with a bounded spatial support is not sufficient in order to fix the sequence of truncated LGGEs. We show that this ultimately means that any UV regularised version of the local charges, such that they can fix the GGE, is equivalent to quasi-local charges with arbitrary support.

Motivated by the issues found in both versions of the local GGE, we finally proceed to consider an ensemble based on the quasi-local charges. 
In Section \ref{quasilocsec} we address the question if the MGGE is equivalent to the QLGGE. In fact, we show that the momentum GGE is naturally expressed as an exponential of a quasi-local operator in the sense of the direct definition we introduced: investigating the locality properties of the latter we show that the coupling between different points is exponentially damped in the distance, with a typical decay length given by the minimum between the pre-quench and post-quench masses.

We finally investigate the amount of information actually needed in order to determine the GGE: an appealing promise of the LGGE was to be able to determine the steady state through a countable set of information (i.e. the expectation value of the local charges). 
Unlike the set of ultra-local charges, which is discrete, the set of quasi-local charges is \emph{continuous}, thus raising the question if the amount of information that is necessary in the GGE construction is also continuous. We show that the answer to this question is negative: knowledge of the expectation value of a countable set of suitably defined quasi-local charges is sufficient in order to completely fix the GGE.

\section{Mass quench in the Klein Gordon field theory}
\label{massquench}

In this and in the following section we describe the class of quenches we choose to consider and address the issue about the validity of the local GGE in its stronger interpretation (\ref{DLGGE}). Instead the problem of the convergence of the truncated local GGE will be addressed in Section \ref{truncGGE}. We will show that DLGGE is not equivalent to the MGGE: in particular we find and quantify a clear discrepancy between the predictions the DLGGE and of the actual correlators in the steady state.

To demonstrate our argument in the simplest possible example, in this section we consider a quantum quench of the mass parameter in the Klein-Gordon field theory. This quench has been studied in \cite{cft2} and the steady state is known in the momentum-space form (\ref{momentumGGE}) so that we can directly test its consistency with DLGGE. We study in parallel the equivalent quench in the lattice version of the model, which is nothing but a system of coupled harmonic oscillators. As we will see, the discrepancy between the ultra-local GGE and the exact steady state is absent in the lattice version, meaning that the problem arises only in the continuum limit.

We consider the Klein-Gordon model in one spatial dimension which is described by the Lagrangian 
\begin{equation}
\mathcal{L}=\int dx \; \frac{1}{2}\partial_{\mu}\phi\partial^{\mu}\phi-\frac{m^2}{2}\phi^2\,\,\, .\label{KG}
\end{equation}
In this simple model, the local charges are quadratic in the fields \cite{muss}:
\begin{equation}
\mathcal{I}_{2n}^{L}=\frac{(-1)^n}{2}\int dx \; \partial_{t}\phi\partial_{x}^{2n}\partial_{t}\phi+\phi\left[\partial_{x}^{2n}(-\partial_{x}^{2}+m^2)\right]\phi=\int \frac{dk}{2\pi}\;k^{2n} E(k)a^\dagger_k a_k\,\,\, ,\label{lch}
\end{equation}
\begin{equation}
\hspace{2pc}\mathcal{I}_{2n+1}^{L}=(-1)^n\int dx\; \partial_{t}\phi\partial_{x}^{2n+1}\phi=\int \frac{dk}{2\pi}\;k^{2n+1}a^\dagger_k a_k\,\,\, .\label{lchbis}
\end{equation}

The system is initialized in the ground state of the model with mass $M$ and then the mass parameter is changed to $m$. The Hamiltonian of the system is readily written as:
\begin{equation}
H= \int dx \; \frac{1}{2}\Pi^2 + \frac{1}{2}(\partial_x\phi)^2+\frac{1}{2}m^2\phi^2 = 
\int \frac{dk}{2 \pi} \;  \frac{1}{2}\tilde{\Pi}_{k}\tilde{\Pi}_{-k}+\frac{1}{2} E(k)^2 \tilde{\phi}_{k} \tilde{\phi}_{-k} = 
\int \frac{dk}{2 \pi} \;   E(k) a^\dagger_k a_k+ \text{const.}\label{B2}
\end{equation}
in standard notation: $\Pi = \partial_t \phi$ is the field conjugate to $\phi$ and  $a_k^\dagger, a_k $ are the creation and annihilation operators with standard bosonic commutation rules $[a_{k},a_{q}^{\dagger}]=2\pi\delta(k-q)$.

The above Hamiltonian admits an obvious lattice discretization in terms of coupled harmonic oscillators:
\begin{equation}
H^{\text{lattice}}=\sum_{j}\frac{1}{2}\Pi_{j}^2+\frac{1}{2\Delta^2}(\phi_{j+1}-\phi_{j})^2+\frac{m^2}{2}\phi_{j}^2 =
\int _{-\pi/\Delta}^{\pi/\Delta}\frac{dk}{2 \pi} \;   E^{\text{lattice}}(k) a^\dagger_k a_k + \text{const.}\,\,\, .\label{B3}
\end{equation}
In the lattice model the integral is constrained in the first Brillouin zone $(-\pi/\Delta,\pi/\Delta)$ where $\Delta$ is the lattice spacing. The dispersion relation in the two cases is:
\begin{equation}
E(k)=\sqrt{k^2+m^2}\hspace{3pc}
E^{\text{lattice}}(k)=\sqrt{2\Delta^{-2}(1-\cos \left(k\Delta\right))+m^2}\,\,\, . \label{B5}
\end{equation}
As shown in \cite{cft2} the relation between the pre-quench creation and annihilation operators $b_{k},b^{\dagger}_{k}$ and the post-quench ones $a_{k},a^{\dagger}_{k}$ is a Bogoliubov transformation
\begin{equation}
a_k=\frac{1}{2}\left(\sqrt{\frac{E(k)}{E_0(k)}}+\sqrt{\frac{E_0(k)}{E(k)}}\right)b_k+\frac{1}{2}\left(\sqrt{\frac{E(k)}{E_0(k)}}-\sqrt{\frac{E_0(k)}{E(k)}}\right)b_{-k}^{\dagger}\,\,\, ,\label{free16}
\end{equation}
where the index ``0" denotes the pre-quench dispersion relation that corresponds to mass $M$. 
The initial state expectation values of the conserved momentum-mode occupation number operators $a^\dagger_k a_k$ is therefore
\begin{equation}
\braket{a^\dagger_k a_k}_{0}=2\pi\delta(0)\frac{1}{4}\frac{\left(E(k)-E_0(k)\right)^{2}}{E_0(k)E(k)}\,\,\, .
\end{equation}
The formally divergent $\delta(0)$ term comes from the commutator $[b_{k},b_{k}^{\dagger}]=2\pi\delta(0)$ and, as standard in quantum field theory, it can be cured by a simple finite volume regularization $2\pi\delta(0)\to L$ \cite{Weinberg}, with $L$ the (infinite) volume of the system. Clearly this regularization results in an extensive expectation value of the Hamiltonian, as it should be. 
For convenience we dispose of the volume factor embedding it in the definition of the density of excitations $n(k)=L^{-1}a^\dagger_k a_k$, in this way:
\begin{equation}
\braket{n(k)}_{0}=\frac{1}{4}\frac{\left(E(k)-E_0(k)\right)^{2}}{E_0(k)E(k)}\,\,\, .\label{new31}
\end{equation}

Using (\ref{new31}) we can immediately understand that a straightforward application of the LGGE recipe is problematic. As a matter of fact the computation of the Lagrange multipliers associated with the local charges passes through the expectation values of the latter, but the defining integrals (\ref{lch}-\ref{lchbis}) are UV divergent, since $\braket{n(k)}_0\sim |k|^{-4}$ for large $k$. 

We postpone the discussion about the delicate issue of fixing the GGE to Section \ref{truncGGE}, since in the current formulation of the LGGE we find an obstacle at a more fundamental level. 
In fact, before asking how we could fix the Lagrange multipliers, we should understand if they actually exist, i.e. if $\braket{n(k)}_0$ is compatible with a DLGGE.

In fact we will show that, without explicitly deriving the Lagrange multipliers, the LGGE form imposes strong analyticity constraints that are sufficient in order to compare it with the exact one as given by the MGGE and to obtain information about the large distance behaviour of correlators. 
From this comparison we can easily see that the LGGE is not equivalent with the exact steady state which is given by the MGGE (\ref{momentumGGE}). 

The excitations densities in the MGGE are:
\begin{equation}
\braket{n(k)}_{MGGE}=\frac{1}{e^{\lambda(k)E(k)}-1}\,\,\, .\label{newsp14}
\end{equation}
Imposing $\braket{n(k)}_0=\braket{n(k)}_{MGGE}$ we compute the Lagrange multipliers or ``effective temperature" $\lambda(k)$
\begin{equation}
\lambda(k)=\frac{1}{E(k)}\log\left(\frac{E_0(k)+E(k)}{E_0(k)-E(k)}\right)^{2}\,\,\, .\label{new32}
\end{equation}
We note that $\lambda(k)$ is an even function of $k$ as it should be, since the initial state is parity invariant. 

A DLGGE (\ref{localGGE}) in the continuum is constructed using an infinite sum of the local charges. Due to parity invariance only the even charges $\mathcal{I}_{2n}^{L}$ defined in (\ref{lch}) enter in this construction. Writing this sum in momentum space we have
\begin{equation}
\sum_{n=0}^{\infty}\lambda_{n}\mathcal{I}_{2n}^{L}=\int \frac{dk}{2\pi}\left(\sum_{n=0}^{\infty}\lambda_{n}k^{2n}\right)E(k)a^\dagger_k a_k+\text{constant}\,\,\, \label{newsp15}
\end{equation}
so that, comparing the DLGGE (\ref{localGGE}) with the MGGE (\ref{momentumGGE}), we identify the DLGGE prediction for $\lambda(k)$ with the power series
\begin{equation}
\lambda_{DLGGE}(k) = \sum_{n=0}^{\infty}\lambda_{n}k^{2n}\,\,\, .\label{newsp15b}
\end{equation}

We conclude that the momentum-dependent effective temperature $\lambda(k)$ for a DLGGE is necessarily defined through a power series, regardless of the actual values of the Lagrange multipliers $\lambda_n$.
 It must be stressed that (\ref{newsp15b}) does not state that the effective temperature is Taylor-expandable around $k=0$, but that it coincides with the series itself: if the latter has a finite convergence radius $R_{conv}$ and diverges beyond it, then the effective temperature becomes infinite as well. 
Conversely, in order for the exact steady state, which is given by the MGGE, to be equivalent to a DLGGE, $\lambda(k)$ must be a power series in terms of $k^2$.  
More importantly, in order for the DLGGE to be correct, this power series must be such that the excitation density computed through the DLGGE converges to the actual value for any real momentum, since otherwise it would lead to incorrect predictions for the values of local observables, as we will show next.
While it is true that the correct effective temperature $\lambda(k)$ (\ref{new32}) can be Taylor expanded around zero, the radius of convergence $R_{conv}$ of this series is $R_{conv}=\min\{m,M\}$, because this is the distance from the origin of the closest of the branch-cut singularities of (\ref{new32}), which are located at $k=\pm im$ and $k=\pm iM$ in the complex plane.
This means that the correct effective temperature (\ref{new32}) cannot be described through a unique power series valid for any real momentum. This may not seem to be a critical problem, since the GGE is supposed to describe correctly only local observables, not global ones like momentum-space quantities: the equivalence between DLGGE and MGGE must be tested on expectation values of local observables. However $\langle n(k)\rangle_0$ is directly linked through Fourier transform to the steady state two-point correlation function $\langle\phi(x,t)\phi(y,t)\rangle_{t\to\infty}$ which \emph{is} a local observable. 
\begin{equation}
\braket{\phi(x)\phi(y)}_{t\to\infty}=\braket{\phi(x)\phi(y)}_{MGGE} 
=\int \frac{dk}{2\pi}\frac{e^{ik(x-y)}}{2\sqrt{k^{2}+m^{2}}} (2\braket{n(k)}_0+1)\,\,\, .\label{2pf-mq}
\end{equation}
In particular, the non-analyticity point of the excitation density $\langle n(k)\rangle_0$ that is closer to the real axis determines the long-distance decay of the two-point correlator. As a result the above discrepancy becomes evident in the long-distance asymptotic behavior, which is what any continuous QFT is expected to capture.

We postpone a detailed comparison between the predictions of DLGGE and MGGE on the two-point correlator to Section \ref{mainsec}, where we discuss the general case of an arbitrary pre-quench theory, but it is worth anticipating the results. In particular, the discrepancy is more evident when the pre-quench mass $M$ is chosen smaller than the post-quench one, $M<m$. Due to the branch cut at $k=\pm i M$ in the excitation density (\ref{new31}), the leading asymptotics behavior of the two point correlator (\ref{2pf-mq}) is an exponential decay $\sim e^{-M|x-y|}$ with multiplicative algebraic corrections. In contrast, a LGGE would predict asymptotes $\sim e^{-m|x-y|}$, if the radius of convergence of (\ref{newsp15b}) is infinite (and a non-exponential decay, otherwise), which is different from the correct one. The detailed derivation of these asymptotes, as well as the discussion of the opposite case $m<M$ and of a finite convergence radius for (\ref{newsp15b}) can be found in Section \ref{mainsec}.

We now study the lattice system: the local charges are a simple discretization of the continuum case and as before only the even charges are needed.
\begin{equation}
(\mathcal{I}_{2n}^{L})^{\text{lattice}}=\frac{1}{2}\sum_{j}\Pi_{n+j}\Pi_{j}+\phi_{n+j}\left(m^2\phi_{j}+\frac{1}{\Delta^2}(2\phi_{j}-\phi_{j+1}-\phi_{j-1})\right)=\int_{-\pi/\Delta}^{\pi/\Delta}\frac{dk}{2\pi}\;\cos(k\Delta n) E^{\text{lattice}}(k)a^\dagger_ka_k\;\label{B8}
\end{equation}
Therefore, a LGGE on the lattice can be written in the momentum space as:
\begin{equation}
\sum_{n}\lambda_{n}^{\text{lattice}}(\mathcal{I}_{2n}^{L})^{\text{lattice}}=\int_{-\pi/\Delta}^{\pi/\Delta} \frac{dk}{2\pi}\left(\sum_{n=0}^{\infty}\lambda_{n}^{\text{lattice}}\cos(k\Delta n)\right)E^{\text{lattice}}(k)a^\dagger_ka_k+\text{constant}\,\,\, .
\end{equation}

In marked contrast to the continuum case, in the lattice we reach the conclusion that in order for the LGGE to be equivalent to the MGGE the effective temperature $\lambda(k)$ must be a \emph{Fourier series} of $k$ instead of a power series. The correct effective temperature is still given by (\ref{new32}) with $E_0(k)$ and $E(k)$ replaced by the lattice dispersion relation. We notice that, since the effective temperature is a smooth function of $k$ over all the Brillouin zone for any $\Delta\ne 0$, it possesses an \emph{absolutely convergent} Fourier series \cite{bifu}. 
This means that, unlike the continuum case, in the lattice the direct definition of the local GGE is equivalent to the correct MGGE, since the latter can be expressed as the exponential of a sum of local charges unambiguously determined by the Lagrange multipliers.

Comparing these two models we can have a better understanding of the continuum limit. Heuristically to pass from the lattice to the continuum we would make the correspondence $\phi_{j}\leftrightarrow\phi(\Delta j)$, then Taylor expand the field when its position is shifted by a finite number of sites $\phi_{j+1}=\phi(\Delta j)+\Delta\partial_{x}\phi(\Delta j)+...$ and find the continuum local charges as linear combinations of those on the lattice. As already pointed out in \cite{Panfil}, this is not the only way to take the continuum limit: we can consider a combined limit for $(\mathcal{I}^L_{2n})^{\text{lattice}}$, letting $\Delta\to 0$ but keeping $n\Delta= y $ constant, thus naturally finding the bosonic analogue of the set of quasilocal charges introduced in \cite{Panfil}.
\begin{equation}
\mathcal{I}^{\text{QL}}( y )=\frac{1}{2}\int dx \; \partial_{t}\phi(x+ y )\partial_{t}\phi(x)+\phi(x+ y )(-\partial_{x}^{2}+m^2)\phi(x)=\int \frac{dk}{2\pi}\; \cos(ky)E(k)a^\dagger_k a_k+\text{const.}\,\,\, .\label{n24}
\end{equation}
We write only the even charges, but of course a similar operation can be performed on the odd ones.
It is straightforward to verify from scratch that the quasilocal charges $\mathcal{I}^{\text{QL}}( y )$ are indeed integrals of motion of the continuum model. Moreover they are complete in the sense that they are sufficient in order to describe the MGGE
\begin{equation}
\int \frac{dk}{2\pi}\; \lambda(k)E(k)a^\dagger_k a_k=\int dy\; \mathcal{K}( y )\mathcal{I}^{\text{QL}}( y ),\hspace{3pc} \text{with }\,\,\mathcal{K}( y )=\int \frac{dk}{2\pi}\lambda(k)\cos(ky)\label{newsp20}
\end{equation} 
and the Fourier transform of $\lambda(k)$ given by (\ref{new32}) is well defined.
Of course, for a quench protocol that breaks parity invariance we must use also the odd charges.

Apart from demonstrating that local charges are insufficient to describe the steady state of the mass quench and that quasilocal charges are instead complete, it is useful to quantify how much nonlocal the GGE is, i.e. how far from local the kernel $\mathcal{K}( y )$ must be in order to describe the correct steady state as given by the MGGE for the mass quench. Obviously considering an arbitrary kernel $\mathcal{K}( y )$ in (\ref{newsp20}) allows for quite nonlocal features, e.g. in the extreme case of a constant kernel points that are infinitely far away would be coupled. Understanding how fast $\mathcal{K}$ decays is important from the viewpoint of a truncation of the GGE \cite{int16}, when we are interested only on observables within a finite subsystem: in practice, we cannot measure experimentally an infinite number of charges, therefore knowing in advance whether or not the contribution of $\mathcal{I}^{\text{QL}}( y )$ is negligible would be very helpful. 

Using the parity invariance of the effective temperature, $\mathcal{K}( y )$ can be computed as:
\begin{equation}
\mathcal{K}( y )=\int \frac{dk}{2\pi}\lambda(k)e^{iky}\,\,\, ,\label{newsp37}
\end{equation}
so that the large distance behavior of $\mathcal{K}( y )$ can be readily estimated from the non-analyticity points of $\lambda(k)$ in the complex plane.
Since $\mathcal{K}( y )$ is symmetric in its argument we can suppose $ y >0$, then we can move the integration path in the upper complex plane, integrating around the non-analyticity points and branch-cuts. 
The non-analyticity point closest to the real axis determines the leading behavior of $\mathcal{K}$ at large distances. In particular, if this point has imaginary part $\xi$, then $\mathcal{K}$ is exponentially damped $\mathcal{K}( y )\sim e^{-\xi| y |}$, in general with extra non-exponential corrections. From the expression of the effective temperature (\ref{new32}) with the relativistic dispersion relation, we see that the first non-analyticity point is a branch-cut singularity at $k$ equal to the smallest mass among $m$ and $M$, therefore:
\begin{equation}
\mathcal{K}( y )\sim e^{-| y |\min(m,M)}\,\,\, .
\end{equation}

Therefore even though all quasilocal charges $\mathcal{I}^{\text{QL}}( y )$ should in principle enter in the construction of the GGE, we have a very natural cutoff in terms of the smallest mass involved. In particular, the larger the masses are, the closer the MGGE or the equivalent QLGGE is to a DLGGE, since only quasilocal charges with a smaller spread are needed.

Apart from the long distance behavior, we can also easily characterize $\mathcal{K}( y )$ at short distances, showing that it is logarithmically divergent. At large momenta $\lambda(k)$ is slowly decaying:
\begin{equation}
\lambda(k)=\frac{1}{|k|}\log\left(\frac{16k^4}{(m^2-M^2)^2}\right)+\mathcal{O}\left(\frac{\log|k|}{|k|^3}\right)\,\,\, .\label{newsp23}
\end{equation}

From this expansion, after simple calculations, we can extract the singular behavior of $\mathcal{K}$ at zero:
\begin{equation}
\mathcal{K}( y )\simeq \frac{4}{\pi}\left[\frac{1}{2}(\log y)^2+(\gamma+\log\beta)\log y\right]+\text{const.},\hspace{2pc} y \to 0\,\,\, ,\label{newsp24}
\end{equation}
where $\gamma$ is the Euler-Mascheroni constant and
\begin{equation}
\beta=\frac{1}{2}\sqrt{|m^2-M^2|}\,\,\, .
\end{equation}

Notice that the leading divergence in (\ref{newsp24}) is \emph{universal} and does not depend on the particular mass shift, provided the latter is not zero. 

In the next section we will generalize these results to the more general case of an interacting-to-free quench: we show that the DLGGE is never sufficient to correctly describe the steady state in continuous field theories, we explain how we can see the discrepancy in correlations of local fields and finally we quantify the non-locality properties of the quasilocal charges that are required for the correct description.

\section{Failure of the DLGGE in interacting-to-free quenches}
\label{mainsec}

In the previous section we showed how the DLGGE is not sufficient to correctly describe the steady state of a mass (free-to-free) quench, but actually we can reach the same conclusion in the more general case of any interacting-to-free quench. We will consider a single bosonic field in one dimension with an arbitrary pre-quench interacting theory:
\begin{equation}
\mathcal{L}_{t<0}=\int dx \hspace{3pt}\frac{1}{2}\left(\partial_{\mu}\phi\partial^{\mu}\phi-m^{2}\phi^{2}\right)-V[\phi]\hspace{1pc}\xrightarrow{t=0}\hspace{1pc}\mathcal{L}_{t>0}=\int dx\hspace{3pt} \frac{1}{2}\left(\partial_{\mu}\phi\partial^{\mu}\phi-m^{2}\phi^{2}\right)\,\,\, .\label{new3}
\end{equation}

Interacting to free quenches in field theory have already been studied \cite{intf3,SotiriadisMartelloni} and it has been shown that, whenever the post-quench Lagrangian is massive ($m\ne 0$) and the initial state satisfies the cluster property, we have equilibration of local observables to the gaussian MGGE (\ref{momentumGGE}).  Here we assume exponential clustering, although the above result is expected to hold even for weaker clustering, as long as equilibration takes place \cite{SotiriadisMartelloni}. If the post-quench theory is massless ($m=0$) the situation is more complicated and the system relaxes to a non gaussian ensemble \cite{Sotiriadis}. 

In this work we will focus only on the massive post-quench case. This quench protocol is parity invariant, therefore we can consider only even charges in the construction of the GGE: in the previous section we showed how DLGGE implies analyticity constraints on the effective temperature $\lambda(k)$. Our strategy is to check whether or not these constraints are consistent with $\braket{n(k)}_{0}=\braket{n(k)}_{DLGGE}$. In the mass quench $\braket{n(k)}_{0}$ was known explicitly thanks to the fact that the pre-quench theory was free. In this more general case we cannot perform this task explicitly, nevertheless the knowledge of the analytic properties of $\braket{n(k)}_{0}$ will be sufficient for our purposes and we can obtain such information using the \emph{K\"all\'en-Lehmann spectral representation} for the correlators of $\phi$ and $\partial_{t}\phi$ \cite{Weinberg}. 

This representation is exact (not perturbative) and relies only on relativistic invariance, providing an integral representation for the two point correlators of the fields $\phi$ and $\partial_{t}\phi$ in terms of a \emph{spectral density} $\rho$:
\begin{equation}
\braket{\tilde{\phi}(k)\tilde{\phi}(-k)}_{0}=\frac{L}{2}\int_{0}^{\infty}d\mu^{2}\frac{\rho(\mu^{2})}{\sqrt{k^{2}+\mu^{2}}};\hspace{2pc}
\braket{\tilde{\Pi}(k)\tilde{\Pi}(-k)}_{0}=\frac{L}{2}\int_{0}^{\infty}d\mu^{2}\rho(\mu^{2})\sqrt{k^{2}+\mu^{2}}\,\,\, .\label{new21}
\end{equation}
Above $\tilde{\phi}(k)$, $\tilde{\Pi}(k)$ are the Fourier transforms of the fields
 and the $L$ factor comes from the usual finite volume regularization of $\delta(0)$ \cite{Weinberg}. 
 
\begin{figure}
\begin{center}
\includegraphics[scale=0.5]{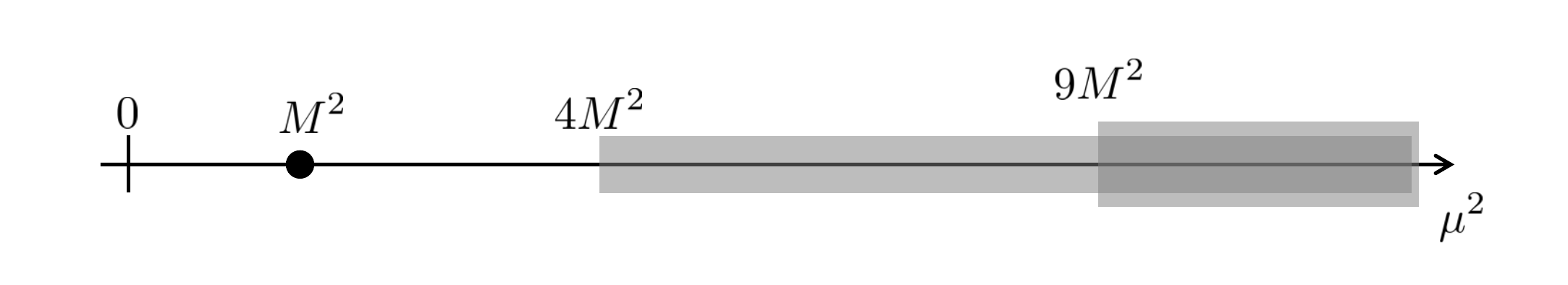}
\caption{Representation of the points and domains where the spectral function $\rho(\mu^{2})$ is nonzero, in the simplest case of only one particle of mass $M$. The spectral function presents a $\delta-$like singularity in correspondence with the one particle state $M^{2}$, then multiple thresholds in correspondence with multiparticle states, therefore starting at $\mu^{2}=n^{2}M^{2}$ with $n=2, 3, 4, ...$ .} \label{fig1}
\end{center}
\end{figure}

The spectral density is real and positive. It exhibits $\delta-$like singularities in correspondence with the masses of the particles of the pre-quench theory and it has thresholds in correspondence with multiparticle states (Figure \ref{fig1}).
\begin{equation}
\rho(\mu^{2})=\sum_{j=1}^{n}Z_{j}\delta(\mu^{2}-M^{2}_{j})+\theta(\mu^{2}-4M_{1}^{2})\sigma(\mu^{2})\,\,\, .\label{KL30}
\end{equation}

Above, $M_{j}$ are the unknown masses of the particles of the pre-quench theory and they are supposed to be ordered $M_{1}<..<M_{n}$, $Z_{j}$ are the field renormalization parameters and they are positive quantities. The continuous threshold starts from the minimal energy of a two particle state, that is $2M_{1}$. Above, $\theta$ is a Heaviside step-function and other thresholds are hidden in the function $\sigma$ in correspondence with multiparticle states of higher mass particles. It is useful to mention \cite{Weinberg} that the spectral density is normalized to $1$:
\begin{equation}
\int_{0}^{\infty}d\mu^{2}\hspace{3pt}\rho(\mu^{2})=1\,\,\, .\label{newnew22}
\end{equation}
Using the parity invariance of $\braket{n(k)}_{0}$ and (\ref{new21}) we get:
\begin{equation}
\braket{n(k)}_{0}=\frac{1}{4}\int d\mu^{2}\; \rho(\mu^{2})\frac{\left(E(k)-\sqrt{k^{2}+\mu^{2}}\right)^{2}}{E(k)\sqrt{k^{2}+\mu^{2}}}\,\,\, .\label{new26}
\end{equation}

It must be stressed that the use of the K\"all\'en-Lehmann spectral decomposition allows us to go beyond perturbation theory and in particular to compare the LGGE and MGGE \emph{directly in the renormalized theory}, that is after the UV divergent constants (masses, field strengths etc.) have been correctly regularized in the limit of infinite UV cut-off. 
Nevertheless notice that from (\ref{new26}) and (\ref{KL30}), the expectation values of the local charges still remain UV divergent, as it happens also in the free case of Section \ref{massquench}, meaning that local charges are not renormalisable by means of standard Renormalization Group theory. This issue is absent in the expectation values of the quasilocal charges which are instead UV finite, exactly as in the mass quench studied in the previous section. Therefore the same problems in the computation of the DLGGE's Lagrange multipliers are still present in this more general quench.

Instead of considering the excitation density, it is more convenient to study the analytic behavior of a slightly different quantity: 
\begin{equation}
F(k)\equiv E(k)[2\braket{n(k)}+1]\,\,\, .
\end{equation}
Besides allowing a more straightforward comparison between the MGGE and DLGGE analyticity predictions, we will see that this quantity is related to the Fourier transform of the two-point field correlator and different analytic behaviors of DLGGE and MGGE will be translated into differences in the large distance behavior of the two point correlator.

Of course, $\braket{n(k)}_{0}=\braket{n(k)}_{DLGGE}$ is equivalent to $F_0(k)=F_{\text{DLGGE}}(k)$: in the following we will show that this equality is in fact impossible.
The expressions of the two quantities above can be derived from (\ref{newsp14}) and (\ref{new26}) and using the canonical commutation rules:
\begin{equation}
F_{0}(k)=\frac{1}{2}\int_{0}^{\infty}d\mu^{2}\hspace{3pt}\rho(\mu^{2})\frac{2k^{2}+m^{2}+\mu^{2}}{\sqrt{k^{2}+\mu^{2}}}\,\,\, ,\label{new34}
\end{equation}
\begin{equation}
F_{DLGGE}(k)=E(k)\left(1+\frac{2}{e^{\lambda_{DLGGE}(k)E(k)}-1}\right)\,\,\, .\label{newnew20}
\end{equation}

Now we can easily study the analytic behavior of $F_{0}(k)$ and then compare it with that of $F_{DLGGE}(k)$:

\begin{figure}
\begin{minipage}[c]{0.45\textwidth}
\includegraphics[scale=0.4]{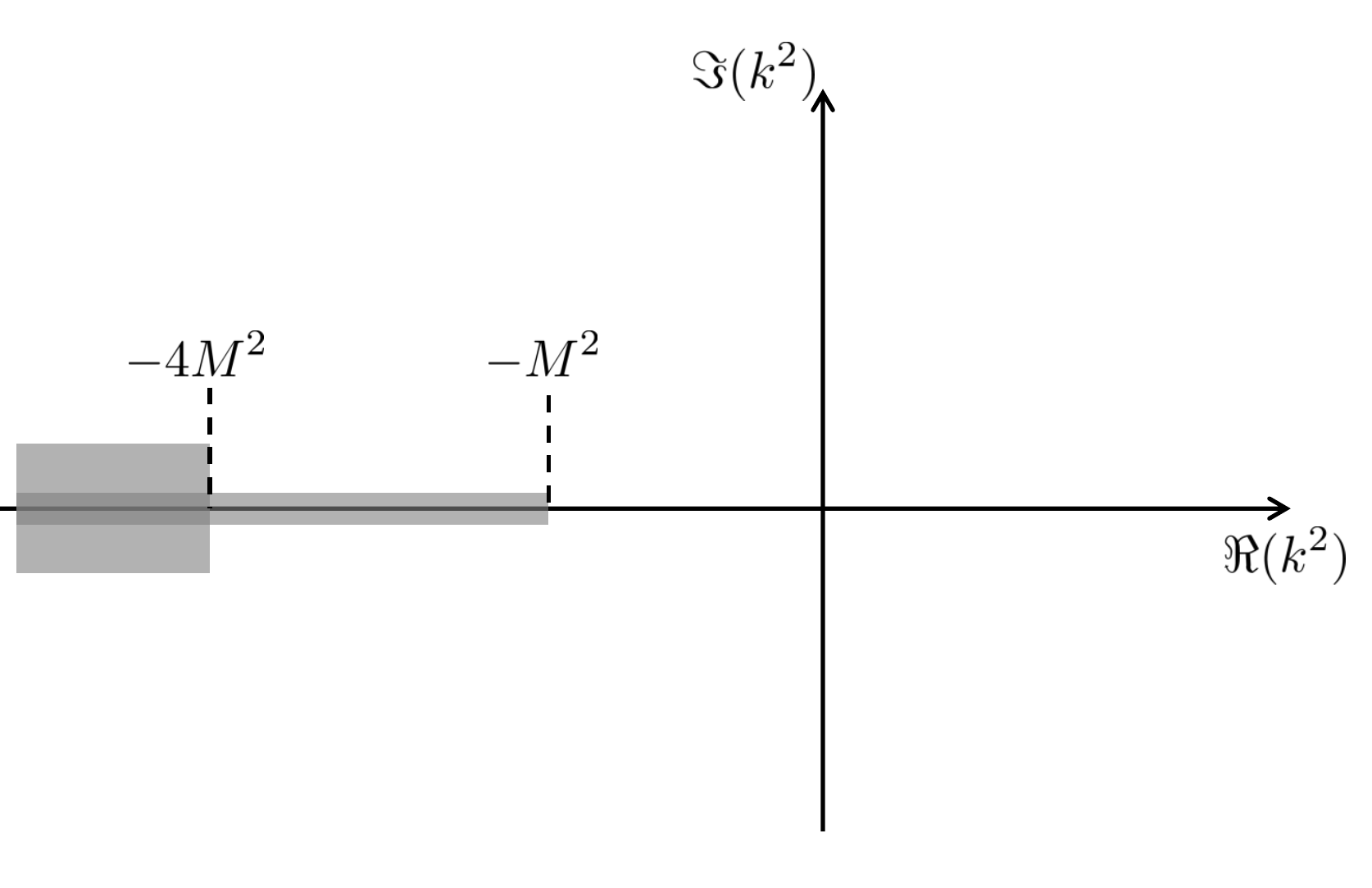}
\caption{Singularities of $F_{0}(k)$ as a function of $k^2$ in the complex plane. For simplicity we assume a pre-quench theory with only one type of particle whose mass is $M$. branch-cuts are located on the negative real axis at $k^2=-(nM)^{2}$ with $n\in\{1,2,3...\}$ corresponding to single or multi-particle excitations that are present in the ground state of the pre-quench theory. Only the branch-cut at $k^2=-M^2$ is a divergent square root singularity, while the others are non-divergent.}\label{fig2}
\end{minipage}
\begin{minipage}[c]{0.07\textwidth}
\hspace{2pc}
\end{minipage}
\begin{minipage}[c]{0.45\textwidth}
\includegraphics[scale=0.4]{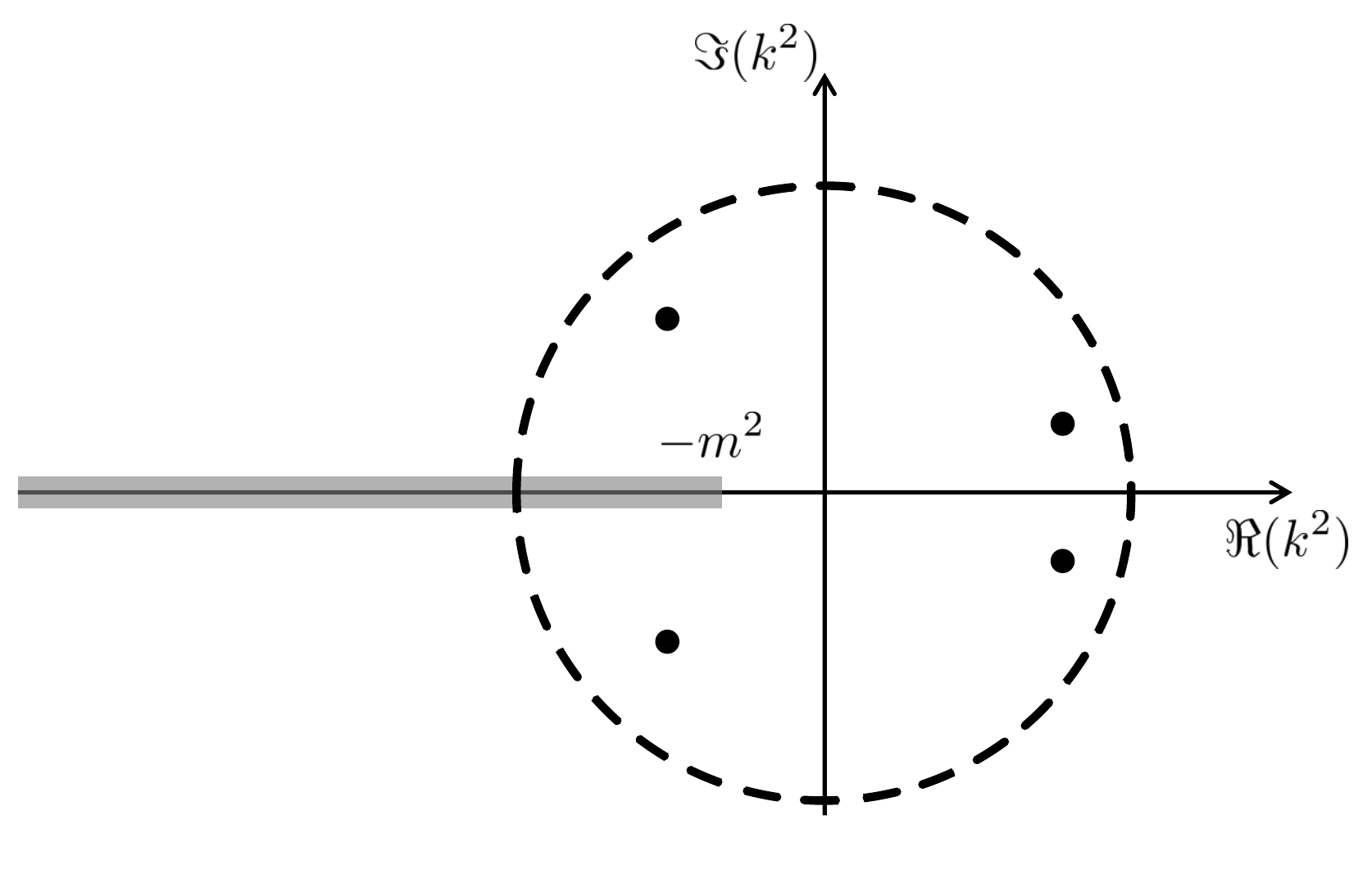}
\caption{Singularities of $F_{DLGGE}(k)$. Inside the disk of radius $R_{conv}$ (depicted by the dashed circle) i.e. the radius of convergence of the series $\lambda_{DLGGE}(k)$, there is always a branch-cut (thick gray line) along the negative real axis starting from $k^{2}=-m^{2}$. Possible additional poles (black dots) are located at the zeros of equation (\ref{new36}). These are the only allowed non-analyticities of $F_{DLGGE}(k)$. \ \\}\label{figpole}
\end{minipage}
\end{figure}
\begin{enumerate}
\item
The analytic structure of $F_{0}(k)$ can be immediately seen from the properties of the spectral density $\rho$ of the pre-quench ground state: we have branch-cut singularities at $k$ equal to the particle masses of the pre-quench theory and also branch-cuts at the thresholds of multiparticle states, as depicted in Figure \ref{fig2}. Notice that all non analyticities in $F_{0}(k)$ are determined by the physical properties of the pre-quench theory. 
From  (\ref{new34}) $F_{0}(k)$ behaves for large momenta as:
\begin{equation}
F_{0}(k)=E(k)\left(1+\frac{1}{k^4}\frac{1}{4}\int d\mu^2\rho(\mu^2)(m^2-\mu^2)^2+\mathcal{O}(1/k^6)\right)\,\,\, .\label{nversion33}
\end{equation}

\item To understand the analytic structure of $F_{DLGGE}(k)$ on the other hand, we use (\ref{newnew20}) and the fact that, as explained in Section \ref{massquench}, $\lambda_{DLGGE}(k)$ must be an analytic function of $k^2$ within the radius of convergence $R_{conv}$ of the series (\ref{newsp15b}): $\lambda_{DLGGE}(k) = \sum_{n=0}^{\infty}\lambda_{n}k^{2n}$. Based on this fact and due to the presence of $E(k)=\sqrt{k^{2}+m^{2}}$, $F_{DLGGE}(k)$ exhibits a branch-cut starting at $k^{2}=-m^{2}$. This brach cut is a non-divergent singularity, unless $\lambda_{DLGGE}(\pm i m)=0$. Note that there is no choice of analytic function $\lambda_{DLGGE}(k)$ for which this branch-cut would be removed. 
Other possible non-analyticity points are located at the zeros of the denominator of (\ref{newnew20}), i.e. the solutions of the equation:
\begin{equation}
\lambda_{DLGGE}(k)E(k)=2ji\pi\, ,\hspace{3pc} j\in \mathbb{Z}\,\,\, . \label{new36}
\end{equation}
Notice that $E(k)$ is analytic everywhere except at $k =\pm i m$. Apart from these points, for any fixed $j$ the corresponding solution $\tilde{k}_j$ is the zero of an analytic function and so it must be an isolated zero. 
This means that the singularities of $F_{DLGGE}(k)$ coming from  zeros of (\ref{new36}) are poles of some finite order. 

These poles and the branch-cut at $k^{2}=-m^{2}$ are the only non-analyticity points of $F_{DLGGE}(k)$ compatible with the locality requirement for the charges included in a DLGGE, as depicted in Figure \ref{figpole}. Notice that changing $\lambda_{DLGGE}(k)$ but keeping it analytic can only affect the position and possibly the order of the poles, but it can neither introduce extra branch-cut singularities, nor remove the branch-cut placed at $k^2=-m^{2}$. 
Note that in order for $\lambda_{DLGGE}$ to describe a physical correlator it must be real for real momenta. Indeed $\braket{\phi(x)\phi(y)}_{\text{DLGGE}}$ must be real which in combination with parity invariance means that its Fourier transform must be real and therefore $\lambda_{DLGGE}$ too. 
This implies that all Lagrange multipliers $\lambda_n$ must be real. Outside of the radius of convergence $R_{conv}$ the series either diverges to $\pm \infty$ or does not exist at all (i.e. the partial sum $\sum_n^{N}\lambda_nk^{2n}$ 
keeps oscillating for increasing $N$). In the second case it is impossible to define (\ref{newnew20}), therefore the only sensible cases are $\lambda_{DLGGE}(k)\to\pm\infty$ for $|k|\ge R_{conv}$. 
Then $F_{DLGGE}(k)=\pm E(k)$ and so neither of these two cases can match with the correct asymptotic behavior (\ref{nversion33}).

\end{enumerate}

It is now clear that a quench protocol like (\ref{new3}) is incompatible with the DLGGE, due to the incompatibility in the analytic structure of $\lambda(k)$.

So far we studied the GGE in the momentum space, but the GGE predictions are meant to be tested on local observables: it is important to understand if the discrepancy we observe for quantities in momentum space has any noticeable consequences in coordinate space. 
To this end we study the two point correlator comparing the DLGGE prediction $\braket{\phi(x)\phi(y)}{}_{DLGGE}$ with the exact large time result $\braket{\phi(x)\phi(y)}{}_{t\to\infty} = \braket{\phi(x)\phi(y)}{}_{MGGE} $. 

We can easily compute the two expressions for $\braket{\phi(x)\phi(y)}$  
in terms of the mode densities:
\begin{equation}
\braket{\phi(x)\phi(y)}_{t\to\infty}=\braket{\phi(x)\phi(y)}_{MGGE} 
=\int \frac{dk}{2\pi}\frac{e^{ik(x-y)}}{2(k^{2}+m^{2})} F_{0}(k)\,\,\, ,
\label{new33a}
\end{equation}
\begin{equation}
\braket{\phi(x)\phi(y)}_{LGGE}
=\int \frac{dk}{2\pi}\frac{e^{ik(x-y)}}{2(k^{2}+m^{2})} F_{DLGGE}(k)\,\,\, .
\label{new33}
\end{equation}

\begin{figure}
\begin{minipage}[c]{0.45\textwidth}
\includegraphics[scale=0.4]{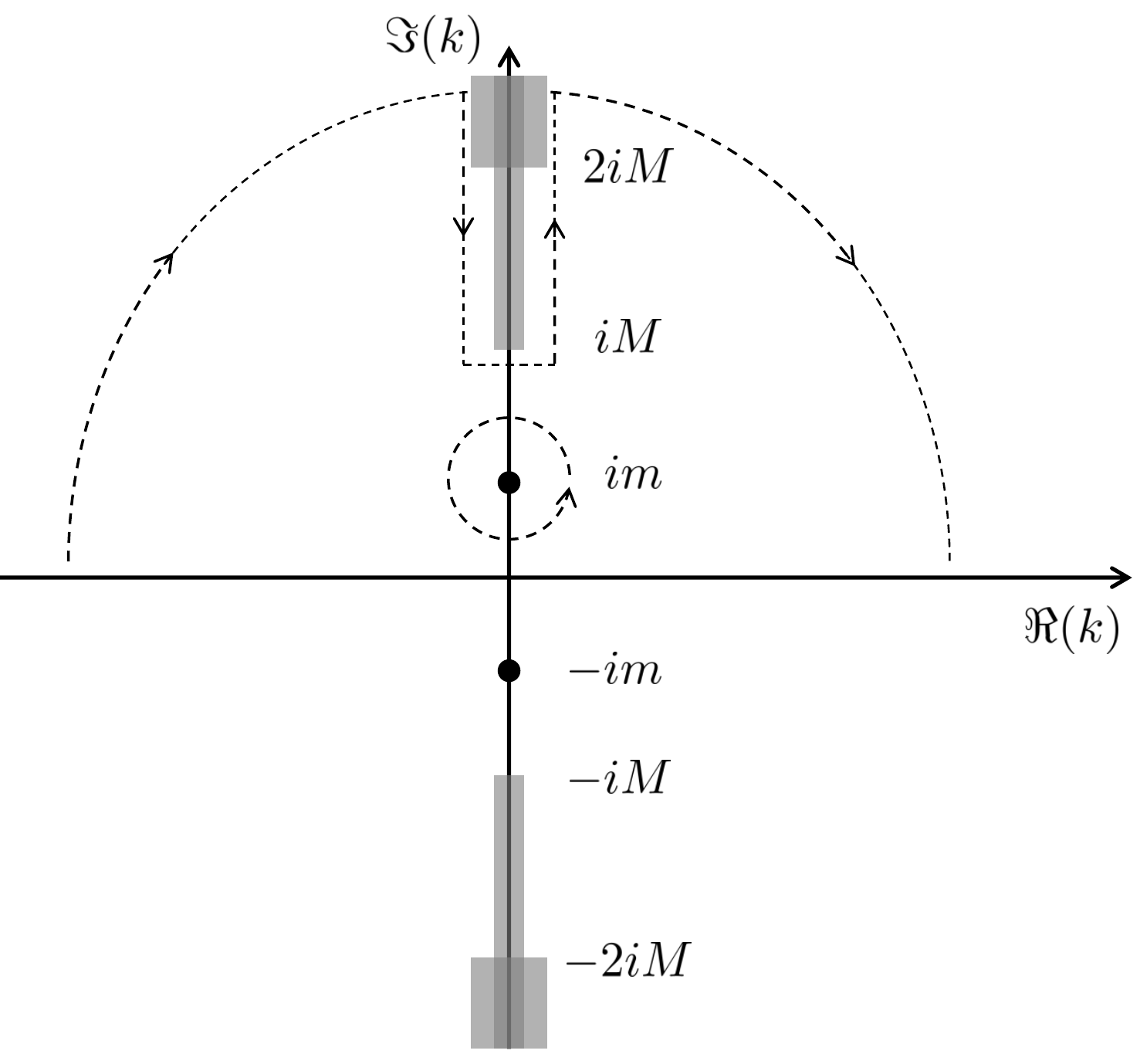}
\caption{Deformation of the integration path for the calculation of (\ref{new33}) for the exact large time limit or equivalently the MGGE expectation value: For simplicity we consider the case of a single type of particle  of mass $M$ in the pre-quench theory. Branch-cuts start at $k=\pm in M$ for $n\in\{1,2,..\}$. First order poles are located at $k=\pm im$, due to the factor $(k^{2}+m^{2})^{-1}$ in (\ref{new33}). \ \\ \ \\ }\label{ldfig2}
\end{minipage}
\begin{minipage}[c]{0.07\textwidth}
\hspace{2pc}
\end{minipage}
\begin{minipage}[c]{0.4\textwidth}
\includegraphics[scale=0.4]{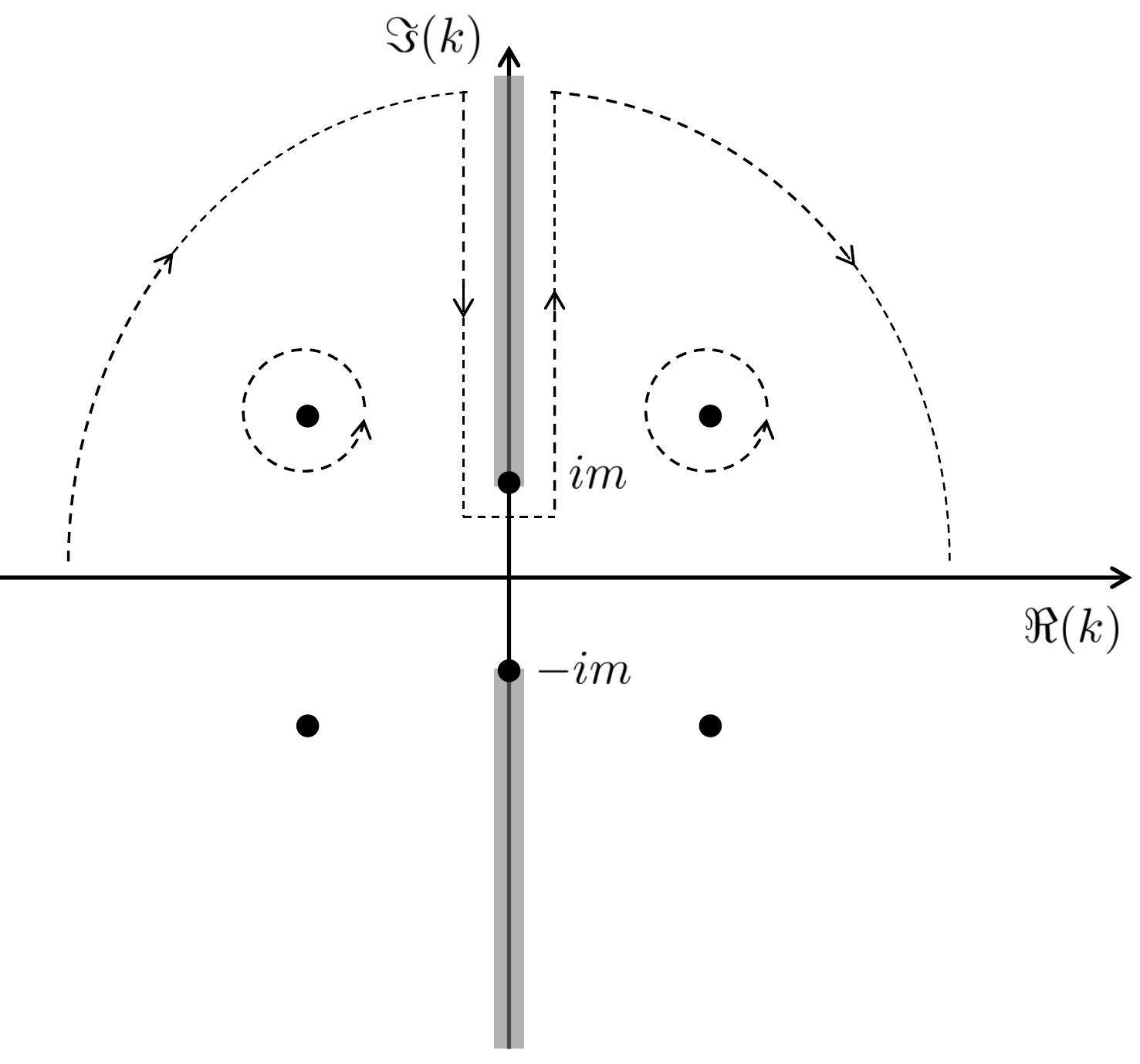}
\caption{Deformation of the integration path for the calculation of (\ref{new33}) for a DLGGE expectation value: Two branch-cuts are present starting from $k=\pm im$. The integrand diverges at the branch-cut points due to the presence of the extra factor of $(k^{2}+m^{2})^{-1}$ in (\ref{new33}). Moreover poles determined by the solutions of (\ref{new36}) may be present symmetrically located with respect to the origin, since (\ref{new36}) is even in $k$.}\label{ldfig1}
\end{minipage}
\end{figure}

Having already showed that there is no way for $F_0$ and $F_{\text{DLGGE}}$ to be equal, we conclude that they cannot give equal two point correlators. This is particularly evident at large distances, where we can easily compare the two predictions showing that they give qualitatively different results.
Since the correlator is symmetric under exchange of $x$ and $y$ we can assume $x>y$ without loss of generality. 
Thanks to the algebraic decay of $F_0(k)$ (\ref{nversion33}), we can compute (\ref{new33a}) and (\ref{new33}) deforming the integration path in the upper complex $k$-plane, going around any non-analyticities of the integrand that are crossed. The different analytic structure between $F_0(k)$ and $F_{DLGGE}(k)$ means different path deformations. Figure \ref{ldfig2} shows the non-analyticities and corresponding path deformation for  (\ref{new33a}) using the K\"all\'en-Lehmann spectral decomposition, while Figure \ref{ldfig1} corresponds to the DLGGE. For the moment we assume that the radius of convergence $R_{conv}$ of $\lambda_{DLGGE}$ is infinite. We will later discuss the case of finite $R_{conv}$. 
To proceed to the calculation of the long distance behavior, we distinguish between two cases: when the post-quench mass $m$ is larger or smaller than the pre-quench mass $M$. 

\begin{enumerate}
\item \textbf{Pre-quench mass smaller than  post-quench mass,} $M<m$:

In this case, the correct long distance behavior which is given by the MGGE (Figure \ref{ldfig2}) is determined by the branch-cut singularity at $k=iM$. Thanks to (\ref{new34}), we know that this is a square root singularity and the long distance behavior can be readily derived:
\begin{equation}
\braket{\phi(x)\phi(y)}_{\text{MGGE}}= \frac{Z}{4\sqrt{2M\pi}} \frac{e^{-M|x-y|}}{\sqrt{|x-y|}}+... \; ,\hspace{3pc}\text{for }|x-y|\to\infty\,\,\, ,\label{new39}
\end{equation}
where $Z$ is the positive field-strength renormalization constant associated with the $\delta$-singularity of $\rho$ at $\mu^2=M^2$ (\ref{KL30}). 
On the other hand, the prediction of a DLGGE (Figure \ref{ldfig1}) cannot be matched with this asymptotic behavior: the branch-cut at $k=im$ results in an exponentially decaying factor $e^{-m|x-y|}$ that is incorrect compared to (\ref{new39}). The only way in which we could capture the correct exponential behavior of (\ref{new39}) would be to choose the effective temperature in such a way that a first order pole appears at $k=iM$, but even then such a pole would give a plain exponential decay and it would never capture the algebraically decaying factor $1/\sqrt{|x-y|}$ of (\ref{new39}). 

\item \textbf{Pre-quench mass larger than  post-quench mass}, $m<M$:

In this case the leading behavior of the exact two point correlator (Figure \ref{ldfig2}) is not given by the branch-cut, but by the pole at $k=im$. Instead, the branch-cut at $k=iM$ gives subleading corrections:
\begin{equation}
\braket{\phi(x)\phi(y)}_{\text{MGGE}}=Ae^{-m|x-y|}+\frac{Z}{4\sqrt{2M\pi}} \frac{e^{-M|x-y|}}{\sqrt{|x-y|}}+...\hspace{3pc}|x-y|\to\infty\,\,\, ,\label{new40}
\end{equation}
where the constant $A$ is:
\begin{equation}
A=\frac{1}{8m}\int d\mu^2 \rho(\mu^2)\sqrt{\mu^2-m^2}\,\,\, .
\end{equation}
Consider now the long distance behavior predicted by a DLGGE (Figure \ref{ldfig1}): this time the branch-cut is in the right position to give the exponentially decaying factor $e^{-m|x-y|}$, but it would not be a pure exponential decay, since we are dealing with a branch-cut and not with a pole. Even if we adjusted the effective temperature to obtain the correct leading behavior, the subleading term will not be exponentially damped $e^{-M|x-y|}$ as predicted by the correct computation (\ref{new40}). In this case the DLGGE fails to catch the correct subleading behavior of the two point correlator at large distances.
\end{enumerate}

Now we consider what happens in the presence of a finite radius of convergence $R_{conv}$. 
As explained earlier, for real $k$ outside of the convergence disk i.e. $|k|>R_{conv}$ the only meaningful possibilities are $\lambda_{DLGGE}(k)=\pm \infty$. 
In this case the two point correlator is:
\begin{equation}
\braket{\phi(x)\phi(y)}_{DLGGE}=\int_{-R_{conv}}^{+R_{conv}} \frac{dk}{2\pi}\frac{e^{ik(x-y)}}{2(k^{2}+m^{2})} \left(F_{DLGGE}(k)\mp E(k)\right)\pm\int_{-\infty}^{+\infty} \frac{dk}{2\pi}\frac{e^{ik(x-y)}}{2\sqrt{k^{2}+m^{2}}}\,\,\, .\label{newnew38}
\end{equation}

\begin{figure}
\begin{center}
\includegraphics[scale=0.4]{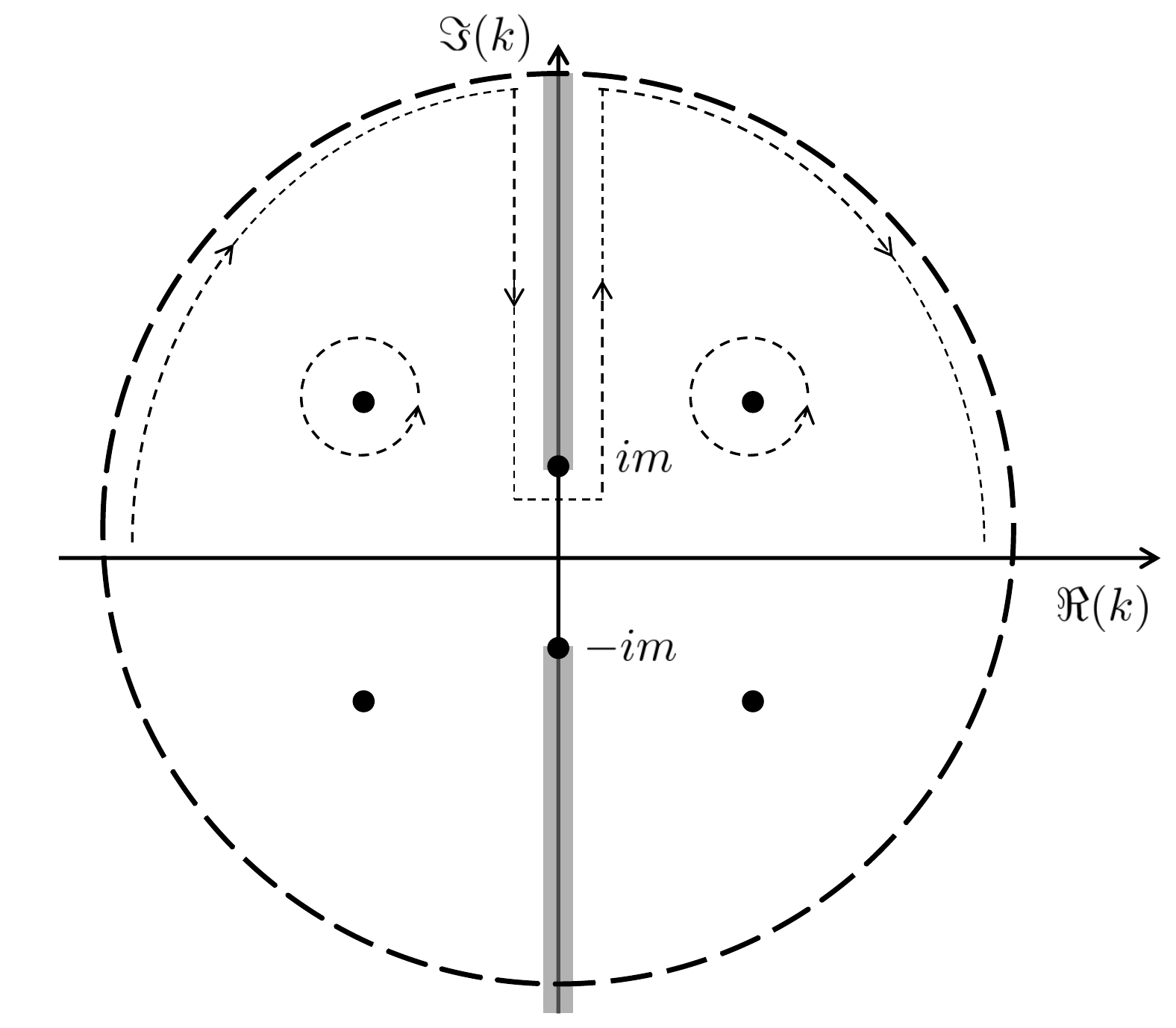}
\caption{Deformation of the integration path in the complex plane for the computation of the two point correlator with a DLGGE, in the case in which the effective temperature has a finite radius of convergence (dashed black line). The interagtion path is constrained in the disk of convergence of the effective temperature.} \label{analyticR}
\end{center}
\end{figure}

The second term, in the large distance limit is exponentially damped $\sim e^{-m|x-y|}$. Consider instead the first term: $\lambda_{DLGGE}$ is well defined and analytic in the whole interval, therefore we can consider the analytic continuation of the integrand to the whole complex plane. 
In particular, $F_{DLGGE}(k)$ is analytic in the whole disk of radius $R_{conv}$: the large distance behavior of the first term can be computed deforming the integration path in the complex plane, enclosing the non-analyticity points, but in contrast to the previous analysis the integration path is constrained inside the disk (Figure \ref{analyticR}).
The leading behavior of the two point correlator is no more exponentially damped: differently from before, we cannot lift the integration path from the real axis and we are forced to touch it at the edges of the disk, therefore the end points of the integration contour give a contribution that decays slower than exponentially.
Actually, if the radius of convergence is taken very large, it could happen that the magnitude of the contributions given by the edges of the integration path is very small, in particular vanishing in the limit $R_{\text{conv}}\to\infty$.
In this perspective, to compute the asymptotic decay of the two point correlator we should rather look at the non analyticity points inside the disk and therefore we come back to the previous discussion.
Overall we conclude that the DLGGE fails not only quantitatively, but also qualitatively, to predict the correct asymptotic behavior for the long distance correlator.

\section{The truncated local GGE} 
\label{truncGGE}

So far we analysed the validity of what we called the direct definition of the LGGE, now we discuss instead the truncated LGGE (\ref{TLGGE}).
The content of this section is organised in two distinct steps. Firstly we address the question if there exists at least one sequence of TLGGE such that (\ref{TLGGE}) is satisfied: this question is well posed independently from the issue of fixing the Lagrange multipliers. 
The second relevant question is which kind of information is needed in order to fix unambiguously the sequence of truncated GGE, i.e. which are the properties of the charges whose expectation value must be known in advance in order to pursue the construction.

While we find a positive answer to the first question through an explicit construction of a suitable sequence of TLGGEs, the second issue is delicate: due to the fact that the expectation values of the local charges are always UV divergent as shown in the previous section, the latter need a proper regularisation.  In the literature this problem has been dealt with by regularising the initial state by means of a sharp or exponential UV cutoff. In this case all ultra-local charges acquire finite values and it is easy to see from (\ref{lch}) and (\ref{2pf-mq}) that the correlation function can be expressed directly in terms of them as a short-distance Taylor expansion. However such a regularisation would modify the initial state in a physically relevant way by introducing an explicit energy scale measurable through the values of physical observables. In fact it is impossible to determine a suitable UV cutoff that would consistently and uniquely reproduce  the short distance observables in the original initial state \cite{boundary}. As we have already pointed out, in equilibrium quantum field theory any UV cutoff has already been consistently removed by the standard renormalisation group procedure and physical observables are independent of them. Thus any additional cutoff would require a new renormalisation scheme and a redefinition of the quench protocol. For these reasons, a GGE that does not invoke any additional cutoff on the field theory is desirable.

To circumvent the problem of UV divergences of the charges, we consider the expectation values of the quasi-local charges $\braket{\mathcal{I}^\text{QL}(y)}$. 
On a qualitative level, the local charges $\mathcal{I}^\text{L}_{2n}$ test the system pointwise and are nothing but derivatives of the quasilocal charges $\mathcal{I}^\text{QL}(y)$ (\ref{n24}) computed at $y=0$. 
Therefore, instead of computing pointwise information, we can consider instead a small interval and address the following question: given an interval $0\le y\le \epsilon$, is the knowledge of $\braket{\mathcal{I}^\text{QL}(y)}$ sufficient in order to determine the GGE?
This question can be seen as a natural regularisation of the local charges that attempts to preserve their locality as much as possible: if $\braket{\mathcal{I}^\text{QL}(y)}$ was regular at $y=0$ (more specifically Taylor expandable) its knowledge in a finite small interval $0\le y\le \epsilon$ would have been equivalent to the knowledge of its Taylor coefficients, i.e. the expectation value of the local charges.
However, as we are going to show, the information contained in any finite interval is simply insufficient to fix the TLGGE, i.e. we ultimately need the knowledge of the expectation value of quasi-local charges at arbitrarily large distances. In particular, this means that any regularisation of the local charges that would be able to fix the GGE, would be essentially equivalent to quasi-local charges with an arbitrary range, consistently with the regularised version of the local charges used in \cite{PME2}.

After this short summary, we start constructing the required sequence of TLGGE: as we have already seen, the use of local charges in the construction of the GGE is encoded in the effective temperature that must be an even power series in momentum. Of course, in the case of a truncated LGGE the series is finite and we obtain a polynomial. In particular, a TLGGE with $N$ charges is associated with a polynomial of degree $N$ in the variable $k^2$
\be
\rho_{TLGGE}^{(N)}\hspace{2pc}\longleftrightarrow\hspace{2pc} \lambda^{(N)}(k)=\sum_{n=0}^N\lambda^{(N)}_n k^{2n}\label{n48}\,\,\, .
\ee

Based on our knowledge about the correctness of the MGGE, the convergence of the truncated  local GGE (\ref{TLGGE}) is translated into the existence of a sequence of polynomials $\lambda^{(N)}(k)$ such that
\be
\int \frac{dk}{2\pi}\frac{e^{ik(x-y)}}{2\sqrt{k^{2}+m^{2}}} (2\braket{n(k)}_0+1)=\lim_{N\to\infty}\int \frac{dk}{2\pi}\frac{e^{ik(x-y)}}{2\sqrt{k^{2}+m^{2}}} (2\braket{n(k)}_{TLGGE}^{(N)}+1)\,\,\, ,\label{n48}
\ee
where
\be
\braket{n(k)}_{TLGGE}^{(N)}=\frac{1}{e^{\lambda^{(N)}(k)E(k)}-1}\,\,\, .\label{49}
\ee

As already anticipated, we now show that such a sequence indeed exists by means of an explicit construction. As a preliminary step, we introduce the polynomials $q_N(k)$ defined as
\be
q_{N}(k)=\frac{1}{\Lambda_N}\sqrt{\frac{N}{2\pi}}\left(1-\left(\frac{k}{2\Lambda_N}\right)^2\right)^{2N}\,\,\, .
\ee

Above, $\Lambda_N$ is a $N-$dependent parameter with dimensions of energy, on which we will comment further in a while.
The polynomials $q_N$ possess some features important for what follows: first of all they are always positive $q_N(k)\ge 0$ for any real momentum. More importantly, consider their asymptotic behaviour for large $N$ in the region $|k|<2\Lambda_N$: if $\Lambda_N$ is chosen in such a way that $N/\Lambda_N^2\to\infty$, then
\be
q_N(k)\simeq\frac{1}{\Lambda_N}\sqrt{\frac{N}{2\pi}}e^{-2N\left(\frac{k}{2\Lambda_N}\right)^2}\,\,\, .
\ee
and the above gaussian is more and more peaked in the $N\to\infty$ limit, until we obtain a Dirac $\delta$-function centered at $k=0$. 
Moreover, if we choose $\Lambda_N$ in such a way that it diverges in the $N\to \infty$ limit, but at the same time $N/\Lambda_N^2\to\infty$, the polynomials $q_N(k)$ tend to a $\delta$-function over an increasing interval of momenta from $-\Lambda_N$ to $+\Lambda_N$. For example, we can require $\Lambda_N\propto N^{1/3}$: the proportionality constant has the dimension of a momentum, but its actual value is completely irrelevant in the following discussion.

From the polynomials $q_N(k)$ and the correct value of the effective temperature $\lambda(k)$, we construct now the effective temperatures for our truncated LGGEs
\be
\lambda^{(2N)}(k)=\int_{-\Lambda_N}^{\Lambda_N}dp\,\, \lambda(p)q_N(k-p)+C\left(\frac{k}{\Lambda_N}\right)^{2N}\label{n52}\,\,\, ,
\ee
where $C$ is a positive constant with the dimension of an inverse energy, but its actual value is completely irrelevant. Since $q_N$ are even polynomials of order $4N$, the $\lambda^{(2N)}(k)$ defined above is an even polynomial of the same order as well, thus it can be interpreted as the effective temperature associated with a TLGGE constructed out of the first $2N$ even local charges. It should be clarified that by this it is not meant that the TLGGE constructed in this way gives correctly the values of the first $2N$ local charges, which is not literally possible due to their UV divergences. Instead, this ensemble should be considered in the sense of the definition of the introduction, i.e. simply as an element of a sequence of TLGGEs designed to converge to the correct steady state.

Notice also that, being the effective temperature $\lambda$ and the polynomials $q_N$ both positive functions, the $\lambda^{(2N)}$ polynomial we defined is strictly positive as well: this is important to avoid a negative or singular excitation density when (\ref{n52}) is plugged in (\ref{49}).
Consider now the behavior of $\lambda^{(2N)}(k)$ in the large $N$ limit: for any $|k|<\Lambda_N$ we get that $\lambda^{(2N)}(k)$ approaches the correct effective temperature $\lambda(k)$. This can be easily seen from the fact that for $|k|<\Lambda_N$ the argument of $q_N(k-p)$ is always in the region where $q_N$ mimics the delta function, since $|k-p|<2\Lambda_N$ over all the integration domain. In the same region, the additional term in (\ref{n52}) does not matter, because for $|k|<\Lambda_N$ it gives an exponentially vanishing contribution. 
Since $\Lambda_N$ is chosen to be divergent in $N$, $\lambda^{(2N)}(k)$ approaches the correct value $\lambda(k)$ on an interval of an increasing, divergent, length.

However, when $\lambda^{(2N)}$ is used in the integral (\ref{n48}), its argument is not  constrained to the interval $|k|<\Lambda_N$, but rather runs over the whole real axis. 
In particular,  the integral over the domain $|k|>\Lambda_N$  (where $\lambda^{(2N)}(k)$ is completely different from $\lambda(k)$) must be guaranteed to give a vanishing contribution when $N\to\infty$: this is guaranteed by the presence of the additional term in (\ref{n52}). Indeed, being the integral of (\ref{n52}) a positive quantity, we have
\be
\lambda^{(2N)}(k)\ge C\left(\frac{k}{\Lambda_N}\right)^{2N}\,\,\, .
\ee

Therefore, for $|k|>\Lambda_N$ the polynomial $\lambda^{(2N)}(k)$ is more and more divergent in $k$ when $N$ is sent to infinity, providing an effective cut off in the consequent excitation energy (\ref{49})
\be
\braket{n(k)}_{TLGGE}^{(2N)}\simeq \braket{n(k)}_0 \theta(\Lambda_N^2-k^2)\, ,\hspace{3pc}N\gg 1\,\,\, ,
\ee
where $\theta$ is the Heaviside Theta function.
This finally guarantees that
\be
\lim_{N\to\infty} \int\frac{dk}{2\pi}\frac{e^{ik(x-y)}}{\sqrt{k^2+m^2}}\braket{n(k)}_{TLGGE}^{(2N)}=\lim_{N\to\infty}\int_{-\Lambda_N}^{\Lambda_N}\frac{dk}{2\pi}\frac{e^{ik(x-y)}}{\sqrt{k^2+m^2}}\braket{n(k)}_{0}=\int_{-\infty}^{+\infty}\frac{dk}{2\pi}\frac{e^{ik(x-y)}}{\sqrt{k^2+m^2}}\braket{n(k)}_{0}\,\,\, .
\ee

The above identity guarantees the convergence of the correlation function that corresponds to the above constructed sequence of TLGGEs to the correct value (\ref{n48}). 

Notice that even though $\Lambda_N$ behaves as an effective cut off, it is introduced through the GGE construction and there is no need to impose any additional external cut off on the field theory itself, thus the above construction is fully compatible with standard RG applied on the initial state.

Even though the required sequence of truncated LGGEs has been constructed, it must be stressed that the previous construction is strongly based on the exact effective temperature $\lambda(k)$, that is ultimately in one-to-one correspondence with the excitation density.
Thus, as we anticipated, we now investigate the amount of information needed to fix unambiguously the sequence of truncated GGEs. In particular, we show that the knowledge of the expectation value $\braket{\mathcal{I}^\text{QL}(y)}$ for $0\le y\le \epsilon$ is insufficient for any finite $\epsilon$.

The proof of this fact is simple: directly from the definition of the quasi-local charges, the expectation value (normalized on the system size $L$) on the initial state is

\be
L^{-1}\braket{\mathcal{I}^{\text{QL}}( y )}_0=\int \frac{dk}{2\pi}\; \cos(ky)E(k)\braket{n(k)}_0\,\,\, .\label{n59}
\ee

As we see, the expectation value of the quasi-local charges is linked to the excitation density by means of a simple Fourier transform.
Of course, the Fourier transform is invertible only if we are allowed to span the whole axis. Thus, the information on the quasi local charges with $0\le y\le\epsilon$ is not enough to unambiguously fix the excitation density. Let then $\braket{n(k)}'_0$ be another excitation density such as 
\be
\braket{n(k)}_0-\braket{n(k)}'_0=\frac{1}{E(k)}\int_{|\tau|>\epsilon} d\tau\cos(\tau k)\eta(\tau)\,\,\, ,
\ee
where $\eta(\tau)$ is any reasonable function, such that both $\braket{n(k)}_0$ and $\braket{n(k)}'_0$ are positive functions for any $k$ (this is needed because any physical relevant excitation density is a positive real function) and such that both give a finite excitation energy. Both $\braket{n(k)}_0$ and $\braket{n(k)}'_0$ are parity invariant $k\to-k$ and when plugged in (\ref{n59}) they give exactly the same result for $0\le y\le \epsilon$, however different excitation densities means different correlators.

Previously we showed how we can construct a sequence of truncated LGGE that reproduces the expectation value of correlators of a given excitation density: therefore, we can construct two sequences of TLGGE, one converging to $\braket{n(k)}_0$ and the other to $\braket{n(k)}'_0$, that give different predictions for the local observables despite having the same expectation value of the quasi-local charges over the selected interval.

Thus, we can conclude that the information contained in an interval of finite length $\epsilon$ is not enough to fix the GGE: we are in principle forced to consider $\epsilon\to\infty$, i.e. quasi local charges with arbitrary spread.

\section{Quasi locality of the GGE}
\label{quasilocsec}

In the previous sections we carefully analyzed the LGGE addressing two inequivalent definitions. In the first case, what we called the direct definition, we show that the local GGE fails to describe the steady state correlators, in particular their asymptotic behavior. In the second notion, i.e. the convergence of the truncated local GGE, we show that it is indeed possible to construct a proper sequence of ensembles that reproduces the correct observables. However a proper identification of the aforementioned sequence ultimately requires the quasi-local charges (\ref{n24}): in this perspective, it is more natural to construct the ensemble directly in terms of the latter. First, as we did for the mass quench, we want to verify that the MGGE can indeed be expressed as a QLGGE. Second, we study the locality properties of the ensemble encoded in the kernel $\mathcal{K}( y )$ (\ref{newsp20}). Lastly, we characterize the amount of information needed to unambiguously fix the QLGGE: despite the continuous nature of the set of quasilocal charges, we show that only a countable set of information is actually needed.

To verify that quasilocal charges correctly describe the exact steady state i.e. the MGGE, we should verify that it is possible to compute the Fourier transform of the effective temperature.
Using the K\"allen-Lehman spectral decomposition (\ref{new26}) we see that $\lambda(k)$ has no singularities for real $k$ and moreover it is easy to see that $\braket{n(k)}_{0}$ decays algebraically for large momenta. We therefore have that for large momenta
\begin{equation}
\lambda(k)\simeq \frac{1}{|k|}\log\left(\frac{k^4}{\beta^4}\right)+\mathcal{O}\left(\frac{\log|k|}{|k|^3}\right), \hspace{3pc} \text{with }
\beta=\left[\frac{1}{16}\int d\mu^2\rho(\mu^2)(m^2-\mu^2)^2\right]^{\frac{1}{4}}\,\,\, .
\end{equation}

This guarantees that the effective temperature has a proper Fourier transform and therefore $\mathcal{K}(y)$ is well defined (\ref{newsp37}) for any $ y \ne 0$, but it has the same  logarithmic divergence in $ y =0$ as in the free mass quench, once we substitute in (\ref{newsp24}) the new definition of $\beta$.
We can now study the decaying properties of $\mathcal{K}$: exactly as we did for the mass quench in Section \ref{massquench}, the non-analyticity point of $\lambda(k)$ closest to the real axis rules the large distance behavior of $\mathcal{K}$.
From (\ref{momentumGGE}) we can express $\lambda(k)$ in terms of $\braket{n(k)}_{0}$:
\begin{equation}
\lambda(k)=\frac{1}{E(k)}\log\left(1+\frac{1}{\braket{n(k)}_{0}}\right)\,\,\, .\label{new19}
\end{equation}

Using (\ref{new26}) and (\ref{new19}), we can readily see that $\lambda(k)$ acquires the non-analyticities of the spectral density, but additional logarithmic branch-cuts may appear too.
However, it is possible to show that in the strip $|\Im(k)|<\min(m,M)$ (where $M$ is the minimum among the pre quench masses) there are no logarithmic branch cuts. Since there is surely a branch cut at $k=i\min(m,M)$ due to the presence of the square roots, this information is enough to extract the long distance behavior of $\mathcal{K}(y)$:
\begin{equation}
\mathcal{K}( y )\sim e^{-\text{min}(m,M)| y |}\,\,\, ,\label{new50}
\end{equation}
where we neglect non exponential corrections.
The inverse masses of the pre-quench and post-quench theory provide the minimum length scale required for an effective truncation of the quasilocal charges that should be included in the QLGGE. Retaining quasi-local charges up to the minimum mass scale is expected to give a good description of the two point correlator, nevertheless any truncation would fail to exactly capture its asymptotic behavior. 

In order to show the absence of logarithmic branch cuts in the strip $|\Im(k)|<\min(m,M)$ we can argue as follows: we choose the logarithmic branch cut of $\log z$ to start from $0$ and to reach $-\infty$ along the real axis, therefore the branch cut is crossed every time the $\Re (z)\le 0$ and $\Im (z)$ changes its sign.
As a first step we can find regions in the complex plane such that the imaginary part of the analytical continuation of the excitation density is different from zero: in these regions there is no logarithmic branch cut.
Looking at (\ref{new19}), this is possible if and only if $\Im\braket{n(k)}_0\ne 0$. Thanks to (\ref{new26}) we have

\begin{equation}
\Im\left[\braket{n(k)}_{0}\right]=\frac{1}{2}\int d\mu^{2}\; \rho(\mu^{2})\;\Im\left[f\left(\frac{2}{|m^2-\mu^2|}\left[k^2+\frac{m^2+\mu^2}{2}\right]\right)\right]\,\,\, ,\label{log1}
\end{equation}
where the complex function $f(z)$ is defined to be
\begin{equation}
f(z)=\frac{z}{\sqrt{z+1}\sqrt{z-1}}\,\,\, .
\end{equation}

Since the spectral density is real and strictly positive, a non vanishing imaginary part of the excitation density is guaranteed as soon as the imaginary part of the integrand has a constant sign for all values of $\mu$ in the support of $\rho$. The imaginary part of $f(z)$ has a constant sign in each quadrant of the complex plane and this allows us to identify regions where the imaginary part of the excitation density is not zero. For example the first quadrant $\Re(z)>0,\Im(z)>0$ in terms of the complex momentum $k$ is identified as:
\begin{equation}
\Re(k)^2-\Im(k)^2+\frac{m^2+\mu^2}{2}>0,\hspace{2pc}\Re(k)\Im(k)>0\,\,\, .
\end{equation}

Solving the first inequality with respect to the imaginary part and requiring the inequality to be satisfied for any $\mu$ in the support of  the spectral density we find a region in the complex plane of the momentum such that the imaginary part of the excitation density is never zero. Considering also the inequalities coming from the other quadrants we have:
\begin{equation}
\Im[\braket{n(k)}_{0}]\ne 0\;\;\forall\; k\in A, \hspace{2pc} A=\left\{ |\Im(k)|<\sqrt{\Re(k)^2+\frac{m^2+M^2}{2}}\; \text{and}\;\Im(k)\Re(k)\ne 0\right\}\,\,\, .\label{log2}
\end{equation}

Because of this condition, if logarithmic branch cuts are present in the strip $|\Im(k)|<\min(m,M)$ they must be located along the imaginary axis. Looking at the expression for the effective temperature (\ref{new26}) it is evident that the integrand along this axis and within the strip $|\Im(k)|<\min(m,M)$ is always real and strictly positive. Since logarithmic branch cuts occur only when the real part of the excitation density $\langle n(k)\rangle_0$ is negative, we conclude that there is no logarithmic branch cut in the strip.

We finally address the problem of quantifying the amount of information needed in order to determine the GGE. Recalling the expression (\ref{n59}) that links the excitation density with the expectation value of the local charges, we understand that the two are linked through a simple Fourier transform. Knowing $\braket{\mathcal{I}^\text{QL}(y)}_0$ for each $y$ we can certainly invert the Fourier transform (\ref{n59}) and obtain the excitation density, which completely fixes the effective temperature and thus the GGE, as already pointed out in \cite{Panfil}.
However, we are going to show that a continuous amount of information is not really needed and we can restrict ourselves to a countable set of suitably defined quasi-local charges. Consider the function $E(k)\braket{n(k)}_0$, whose Fourier transform is the expectation value of the quasi-local charges (\ref{n24}): as it is clear from the K\"all\'en-Lehmann spectral decomposition, this is a positive, bounded, continuous function of $k$.
Moreover, in all the quenches we considered so far and in any meaningful quench, the total energy density is finite
\be
\int\frac{dk}{2\pi} E(k)\braket{n(k)}_0<\infty\,\,\, .
\ee

The above statement, combined with the fact that $E(k)\braket{n(k)}_0$ is positive and bounded, implies that $E(k)\braket{n(k)}_0$ is square-integrable 
\be
\int\frac{dk}{2\pi} |E(k)\braket{n(k)}_0|^2<\infty\,\,\, .
\ee

Therefore, $E(k)\braket{n(k)}_0$ as well as its Fourier transform, belongs to the Hilbert space of the square-integrable functions, that is a well-known separable Hilbert space, i.e. it possesses a countable orthonormal basis that, without loss of generality, can be chosen to be real. In what follows we can use essentially any basis we prefer, but as a concrete example we choose the eigenfunctions of the harmonic oscillator
\be
f_{n}(y)=\frac{(\omega/\pi)^{1/4}}{\sqrt{2^n n!}}e^{-\frac{\omega}{2} y^2}H_n(\sqrt{\omega}y)\label{n70}
\ee
where in the above $\omega$ is a parameter inserted for dimensional reasons, but its actual value does not affect the subsequent discussion, while $H_n(y)$ are the Hermite polynomials.
Then, it is true that we can expand $\braket{\mathcal{I}^\text{QL}(y)}_0$ in terms of these eigenfunctions. Since $\braket{\mathcal{I}^\text{QL}(y)}_0$ is even, only the even functions $f_n$ will enter in its expansion
\be
\braket{\mathcal{I}^\text{QL}(y)}_0=\sum_{n=0}^\infty\, c_{2n}\, f_{2n}(y), \hspace{4pc}c_{2n}=\int dy\, f_{2n}(y)\braket{\mathcal{I}^\text{QL}(y)}_0\,\,\, .
\ee

Therefore the knowledge of the $c_{2n}$ constants determines $\braket{\mathcal{I}^{\text{QL}}(y)}_0$ and then, through Fourier transform, we can fix $E(k)\braket{n(k)}_0$: these equalities hold in the sense of the $\text{L}^2(\text{R})$ norm, but this determines an unique continuous function $\braket{n(k)}_0$.
The excitation density of course fixes the effective temperature and thus the QLGGE.

Therefore, the QLGGE is in one-to-one correspondence with the set of the $c_{2n}$ coefficients, but the latter can be easily seen to be the expectation value of properly defined quasi-local charges:
\be
c_{2n}=\braket{\mathcal{I}^\text{QL}_{2n}}_0,\hspace{4pc}\mathcal{I}^\text{QL}_{2n}=\int dy\, f_{2n}(y)\mathcal{I}^\text{QL}(y)\,\,\, .
\ee

The charges $\mathcal{I}^\text{QL}_{2n}$ are clearly quasi-local: the coupling among different points is given by the kernel $f_{2n}$ that, because of our choice (\ref{n70}), is damped by a gaussian kernel.
Thus, we conclude that the knowledge of the expectation value of the countable set of quasi-local charges $\mathcal{I}_{2n}^\text{QL}$ unambiguously fixes the GGE.

\section{Conclusions}
\label{conclusions}
In this paper we analysed in detail the locality properties of the GGE for interacting-to-free quenches, in the framework of relativistic quantum field theories. The main results and conclusions are summarised below:

\begin{enumerate}
\item Firstly we considered the GGE protocol based on ultra-local charges. In particular, we tested two inequivalent interpretations of the local GGE protocol, one less constrictive than the other. In the first interpretation we required the local GGE to be defined, once for all, as the exponential of an operator constructed in terms of the local charges. We show that this definition does not capture the correct steady state for any initial state, failing both on a quantitative and on a qualitative level. This can be seen on the two-point correlation function and most importantly in its long distance behavior, which is precisely the local observable that quantum field theories are expected to describe. The second definition of the local GGE protocol we tested is in terms of a sequence of truncated local GGEs, that is ultimately required to recover the correct expectation values of local correlators when the number of the used local charges tends to infinity. In this case we found that it is indeed possible to construct the proper sequence of truncated GGEs, but unambiguously determining the latter in the end requires the use of charges with spatial support in arbitrarily large distances, i.e. quasi-local charges.

\item While the momentum space version of the GGE protocol is always correct, looking at it in the coordinate space ultimately requires quasi-local charges. Despite their nonlocal nature, their weight in the GGE density matrix decays exponentially with the distance over a scale of the order of the inverse of the minimum mass involved. This means that a GGE that describes correctly all local observables in a finite subsystem is only mildly nonlocal: the contribution of quasilocal charges outside of the subsystem is exponentially suppressed.
We also addressed the issue of the actual amount of information needed to construct the GGE: despite the continuum nature of the set of quasi-local charges, we showed that the knowledge of the expectation value of a countable set of properly defined quasi-local charges completely determines the GGE.

\item The pathological behaviour of the local GGE naturally appears in continuous models.
 In fact the problem originates in a naive application of the continuum limit, as already pointed out in \cite{Panfil}. A consistent generalisation of the lattice version of the local charges in the continuum limit are the quasilocal charges, instead of the standard local charges of quantum field theory.

\end{enumerate}

\section{Acknowledgement}
We are grateful to Pasquale Calabrese for his suggestions on this work. We also thank Giuseppe Mussardo and Fabian Essler for useful discussions. Lastly we thank the anonymous referee for suggesting the investigation regarding the truncated local GGE. SS acknowledges support of the A*MIDEX project Hypathie (no. ANR-11-IDEX-0001-02) funded by the ``Investissements d'Avenir'' French Government program, managed by the French National Research Agency (ANR).

\end{document}